\PassOptionsToPackage{unicode}{hyperref}
\PassOptionsToPackage{hyphens}{url}
\PassOptionsToPackage{dvipsnames,svgnames,x11names}{xcolor}
\documentclass[
]{article}
\usepackage{amsmath,amssymb}
\usepackage{lmodern}
\usepackage{iftex}
\ifPDFTeX
  \usepackage[T1]{fontenc}
  \usepackage[utf8]{inputenc}
  \usepackage{textcomp} 
\else 
  \usepackage{unicode-math}
  \defaultfontfeatures{Scale=MatchLowercase}
  \defaultfontfeatures[\rmfamily]{Ligatures=TeX,Scale=1}
\fi
\IfFileExists{upquote.sty}{\usepackage{upquote}}{}
\IfFileExists{microtype.sty}{
  \usepackage[]{microtype}
  \UseMicrotypeSet[protrusion]{basicmath} 
}{}
\makeatletter
\@ifundefined{KOMAClassName}{
  \IfFileExists{parskip.sty}{%
    \usepackage{parskip}
  }{
    \setlength{\parindent}{0pt}
    \setlength{\parskip}{6pt plus 2pt minus 1pt}}
}{
  \KOMAoptions{parskip=half}}
\makeatother
\usepackage{xcolor}
\usepackage{color}
\usepackage{fancyvrb}

\DefineVerbatimEnvironment{Highlighting}{Verbatim}{commandchars=\\\{\}}
\newenvironment{Shaded}{}{}

\newcommand{\DecValTok}[1]{\textcolor[rgb]{0.25,0.63,0.44}{#1}}

\newcommand{\FloatTok}[1]{\textcolor[rgb]{0.25,0.63,0.44}{#1}}

\newcommand{\ImportTok}[1]{\textcolor[rgb]{0.00,0.50,0.00}{\textbf{#1}}}

\newcommand{\NormalTok}[1]{#1}
\newcommand{\OperatorTok}[1]{\textcolor[rgb]{0.40,0.40,0.40}{#1}}

\newcommand{\StringTok}[1]{\textcolor[rgb]{0.25,0.44,0.63}{#1}}

\providecommand{\tightlist}{%
  \setlength{\itemsep}{0pt}\setlength{\parskip}{0pt}}
\newlength{\cslhangindent}
\newlength{\csllabelwidth}
\newlength{\cslentryspacingunit} 
\newenvironment{CSLReferences}[2] 
 {
  \setlength{\parindent}{0pt}
  \ifodd #1
  \let\oldpar\par
  \def\par{\hangindent=\cslhangindent\oldpar}
  \fi
  \setlength{\parskip}{#2\cslentryspacingunit}
 }%
 {}
\usepackage{calc}

\ifLuaTeX
\usepackage[bidi=basic]{babel}
\else
\usepackage[bidi=default]{babel}
\fi
\babelprovide[main,import]{american}

\def\languageshorthands#1{}
\ifLuaTeX
  \usepackage{selnolig}  
\fi
\IfFileExists{bookmark.sty}{\usepackage{bookmark}}{\usepackage{hyperref}}
\IfFileExists{xurl.sty}{\usepackage{xurl}}{} 
\hypersetup{
  pdftitle={planetMagFields: A Python package for analyzing and plotting
planetary magnetic field data},
  pdfauthor={Ankit Barik, Regupathi Angappan},
  pdflang={en-US},
  colorlinks=true,
  linkcolor={Maroon},
  filecolor={Maroon},
  citecolor={Blue},
  urlcolor={Blue},
  pdfcreator={LaTeX via pandoc}}

\title{planetMagFields: A Python package for analyzing and plotting
planetary magnetic field data}

\usepackage[affil-it]{authblk}
\usepackage{orcidlink}
\author[1%
  \ensuremath\mathparagraph]{Ankit Barik%
    \,\orcidlink{0000-0001-5747-669X}\,%
    }
\author[1%
  ]{Regupathi Angappan%
    \,\orcidlink{0000-0002-6258-0659}\,%
    }

\affil[1]{Johns Hopkins University}
\affil[$\mathparagraph$]{Corresponding author}
\date{22 February 2024}

\begin{document}
\maketitle

\hypertarget{summary}{%
\section{Summary}\label{summary}}

Long term observations and space missions have generated a wealth of
data on the magnetic fields of the Earth and other solar system planets
(\protect\hyperlink{ref-IGRF13}{Alken et al., 2021};
\protect\hyperlink{ref-Anderson2012}{Anderson et al., 2012};
\protect\hyperlink{ref-Cao2020}{Cao et al., 2020};
\protect\hyperlink{ref-Connerney1987}{Connerney et al., 1987},
\protect\hyperlink{ref-Connerney1991}{1991},
\protect\hyperlink{ref-Connerney2022}{2022};
\protect\hyperlink{ref-Kivelson2002}{Kivelson et al., 2002}).
\texttt{planetMagfields} is a Python package designed to have all the
planetary magnetic field data currently available in one place and to
provide an easy interface to access the data. \texttt{planetMagfields}
focuses on planetary bodies that generate their own magnetic field,
namely Mercury, Earth, Jupiter, Saturn, Uranus, Neptune and Ganymede.
\texttt{planetMagfields} provides functions to compute as well as plot
the magnetic field on the planetary surface or at a distance above or
under the surface. It also provides functions to filter out the field to
large or small scales as well as to produce \texttt{.vts} files to
visualize the field in 3D using Paraview
(\protect\hyperlink{ref-Ahrens2005}{Ahrens et al., 2005};
\protect\hyperlink{ref-Ayachit2015}{Ayachit, 2015}), VisIt
(\protect\hyperlink{ref-visit}{Childs et al., 2012}) or similar
rendering software. Lastly, the \texttt{planetMagfields} repository also
provides a Jupyter notebook for easy interactive visualizations.

\hypertarget{statement-of-need}{%
\section{Statement of need}\label{statement-of-need}}

Planetary scientists studying the magnetic field of planets need to
constantly access, visualize, analyze and extrapolate magnetic field
data. In addition, with technological advancements in space exploration
and planetary missions, we are constantly getting new data for planetary
magnetic fields and hence, better field models. Though reviews of these
field models are often written
(\protect\hyperlink{ref-Schubert2011}{Schubert \& Soderlund, 2011};
\protect\hyperlink{ref-Stanley2014}{Stanley, 2014}), there is very
little software available that provides easy access to these models with
a high level language and a way to easily visualize and analyze them. To
the knowledge of the authors, there are a few publicly available
repositories that are capable of providing access to planetary magnetic
field data and tools to analyze them such as \texttt{JupiterMag}
(\protect\hyperlink{ref-James2024}{James et al., 2024};
\protect\hyperlink{ref-Wilson2023}{Wilson et al., 2023}), \texttt{KMAG}
(\protect\hyperlink{ref-Khurana2020}{Khurana, 2020}),
\texttt{ChaosMagPy} (\protect\hyperlink{ref-Kloss2024}{Kloss, 2024}),
\texttt{SHTools} (\protect\hyperlink{ref-Wieczorek2018}{Wieczorek \&
Meschede, 2018}), \texttt{PlanetMag}
(\protect\hyperlink{ref-Styczinski2024}{Styczinski \& Cochrane, 2024})
and \texttt{libinteralfield}
(\url{https://github.com/mattkjames7/libinternalfield}). Out of these,
only \texttt{libinteralfield} provides data and software to analyze and
access magnetic fields of all planets. However, it is a \texttt{C++}
library which needs to be interfaced with something at a higher level to
enable fast analyses and visualization. Thus, a software package that
has different magnetic field models for all different planets of the
solar system in one place, as well as provides a high level API to
access, analyze and visualize them is not available.
\texttt{planetMagfields} is intended not only to currently fill this
gap, but also to provide a central repository, to be constantly updated,
as more magnetic field models become available.

In addition to the research aspect of our software, the interactive
Jupyter notebook serves as a valuable educational resource, fostering a
deeper appreciation for the complexities of planetary magnetic
environments.

\hypertarget{mathematics}{%
\section{Mathematics}\label{mathematics}}

Magnetic fields in planets are generated by electric currents in a fluid
region inside them through a process called dynamo action
(\protect\hyperlink{ref-Jones2011}{Jones, 2011};
\protect\hyperlink{ref-Schubert2011}{Schubert \& Soderlund, 2011};
\protect\hyperlink{ref-Stanley2014}{Stanley, 2014}). Outside this
region, in the absence of current sources, the magnetic field
\(\vec{B}\) can be written as the gradient of a scalar potential,
\(\vec{B} = -\nabla V\). The potential \(V\) is usually written as an
expansion in orthogonal functions in spherical coordinates
\((r,\theta,\phi)\),

\begin{equation}
  V = R_p \sum_{l,m} \left(\frac{R_p}{r}\right)^{l+1} [g_l^m \cos(m\phi) + h_l^m \sin(m\phi)] P_l^m (\cos\theta)\, ,
  \label{eq:gaussCoeff}
\end{equation}

where, \(g_l^m\) and \(h_l^m\) are called the Gauss coefficients.
\(R_p\) represents the radius of the planet and \(P_l^m\) are associated
Legendre functions of order \(l\) and degree \(m\), where \(l\) and
\(m\) are integers. The above equation can be recast in terms of
spherical harmonics, which is what the code uses.

The raw data obtained from satellites or space missions are usually
inverted to obtain these Gauss coefficients which are the key to
describing the surface magnetic field of a planet as well as how that
field looks at a certain altitude from the surface. The magnetic energy
content on the surface in a certain degree \(l\) is given by the Lowes
spectrum:

\[R_{l} = (l + 1) \sum_{m}\left( \left(g_l^m\right)^2 + \left(h_l^m\right)^2\right),\]

\(l\) plays the role of a wavenumber. Low degrees represent large
spatial features in the field while high degrees represent small scale
features. The maximum available degree \(l_{max}\) of data for a
particular planet depends on the quality of observations.

\hypertarget{benchmarking}{%
\section{Benchmarking}\label{benchmarking}}

\begin{figure}
\centering
\includegraphics[width=0.8\textwidth]{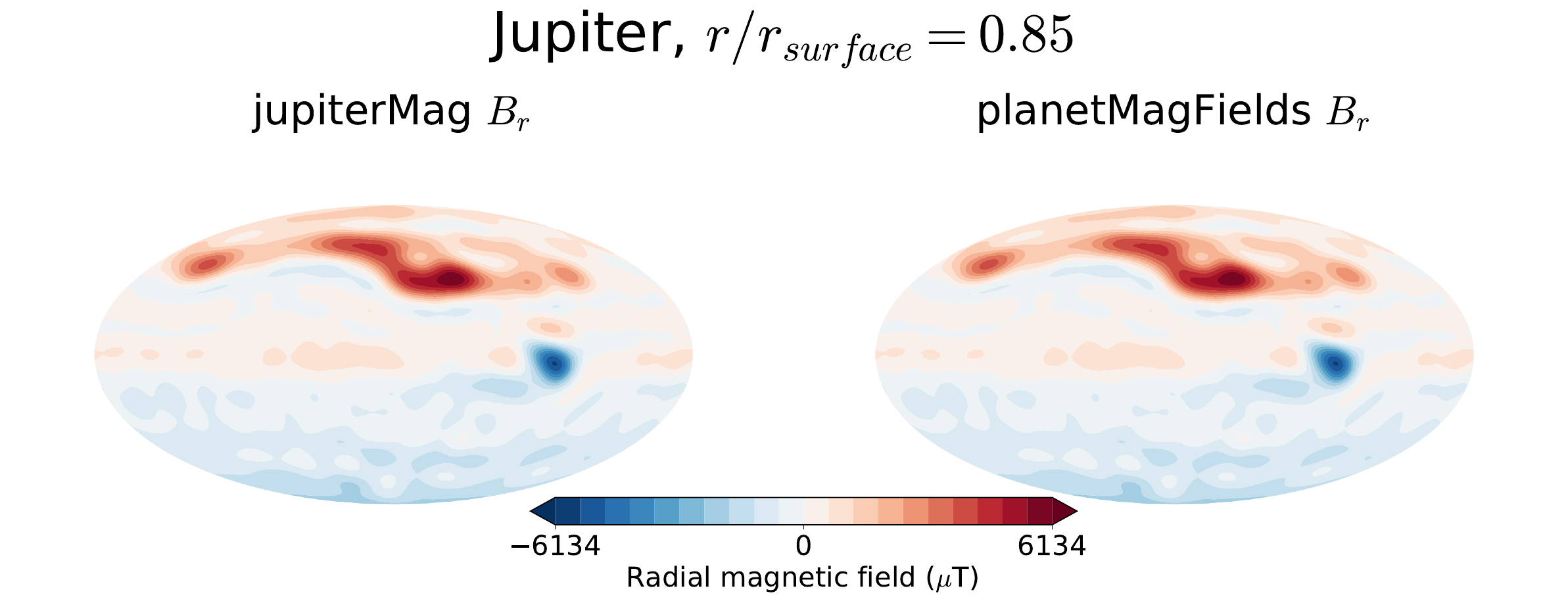}
\caption{Benchmarking the code against publicly available repositories.}
\label{fig:bench}
\end{figure}

We benchmarked our software against two publicly available repositories
: \texttt{JupiterMag} (\protect\hyperlink{ref-James2024}{James et al.,
2024}; \protect\hyperlink{ref-Wilson2023}{Wilson et al., 2023}) for
Jupiter and the \texttt{CHAOS-7}
(\protect\hyperlink{ref-Finlay2020}{Finlay et al., 2020};
\protect\hyperlink{ref-Kloss2024}{Kloss, 2024}) for Earth. For Jupiter,
we compare the field at a depth of 85\% of planetary radius, thus
testing our extrapolation capability while for Earth, we compare the
field on the surface in 2016, testing our implementation of taking into
account changes in the Earth's field in a linear fashion (as is done for
the IGRF model, \protect\hyperlink{ref-IGRF13}{Alken et al., 2021}) The
comparison for Jupiter is shown in Figure \ref{fig:bench}. We also use
these cases in our unit testing.

\hypertarget{description-of-the-software}{%
\section{Description of the
software}\label{description-of-the-software}}

\hypertarget{the-software-package}{%
\subsection{The software package}\label{the-software-package}}

\texttt{planetMagfields} has data files containing Gauss coefficients
from various inversion studies of planetary magnetic models for
different planets. These coefficients are then used to obtain the
magnetic field on a grid of latitude and longitude using equation
\eqref{eq:gaussCoeff}. The main way of accessing the data is through the
\texttt{Planet} class. An example is provided below using IPython
(\protect\hyperlink{ref-ipython}{Pérez \& Granger, 2007}),

\begin{Shaded}
\begin{Highlighting}[]
\NormalTok{ In [}\DecValTok{1}\NormalTok{]: }\ImportTok{from}\NormalTok{ planetmagfields }\ImportTok{import} \OperatorTok{*}

\NormalTok{ In [}\DecValTok{2}\NormalTok{]: p }\OperatorTok{=}\NormalTok{ Planet(name}\OperatorTok{=}\StringTok{\textquotesingle{}jupiter\textquotesingle{}}\NormalTok{,model}\OperatorTok{=}\StringTok{\textquotesingle{}jrm09\textquotesingle{}}\NormalTok{)}
\NormalTok{ Planet: Jupiter}
\NormalTok{ Model: jrm09}
\NormalTok{ l\_max }\OperatorTok{=} \DecValTok{10}
\NormalTok{ Dipole tilt (degrees) }\OperatorTok{=} \FloatTok{10.307870}

\NormalTok{ In [}\DecValTok{3}\NormalTok{]: p.plot(r}\OperatorTok{=}\FloatTok{0.85}\NormalTok{,proj}\OperatorTok{=}\StringTok{\textquotesingle{}Mollweide\textquotesingle{}}\NormalTok{)}
\end{Highlighting}
\end{Shaded}

\begin{figure}
  \centering
  \includegraphics[width=0.5\textwidth]{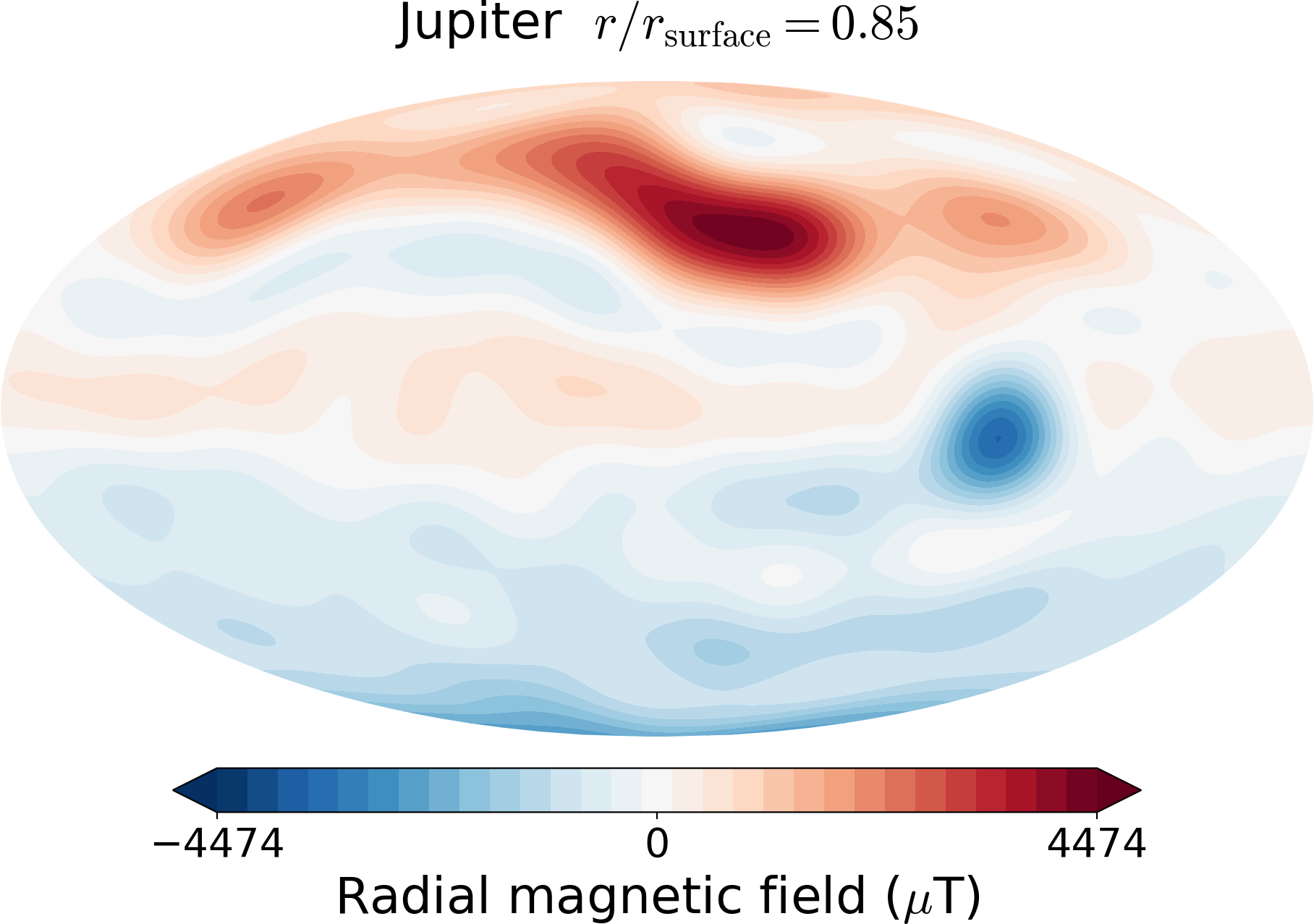}
  \caption[]{Plotting example of Jupiter's radial magnetic field at a depth of 85\% of the planetary radius.}
  \label{fig:jup85}
\end{figure}

The last plot statement produces Figure \ref{fig:jup85} which is the
radial magnetic field at 85\% of the planetary radius. This can be
compared against Figure 1h of Moore et al.
(\protect\hyperlink{ref-Moore2018}{2018}). \texttt{planetMagfields}
primarily uses \texttt{NumPy} (\protect\hyperlink{ref-numpy}{Harris et
al., 2020}), \texttt{Matplotlib}
(\protect\hyperlink{ref-matplotlib}{Hunter, 2007}) and \texttt{SciPy}
(\protect\hyperlink{ref-scipy}{Virtanen et al., 2020}) for most of its
analyses. Further support for various map projections is added through
Cartopy (\protect\hyperlink{ref-cartopy}{Met Office, 2010 - 2015}).
\texttt{planetMagfields} also provides functions to extrapolate and
obtain all components of the magnetic field at a certain depth or height
through spherical harmonic transforms using the SHTns library
(\protect\hyperlink{ref-shtns}{Schaeffer, 2013}). Finally, this
extrapolation also allows one to visualize the field in 3D. To enable
that, \texttt{planetMagfields} uses the \texttt{PyEVTK} library
(\url{https://github.com/paulo-herrera/PyEVTK}) to write \texttt{.vts}
files which can be visualized using software like Paraview or VisIt. An
example for Jupiter is provided below in Figure \ref{fig:jup3d}. A full
list of available features is provided in the documentation.

\begin{figure}
\centering
\includegraphics[width=0.5\textwidth]{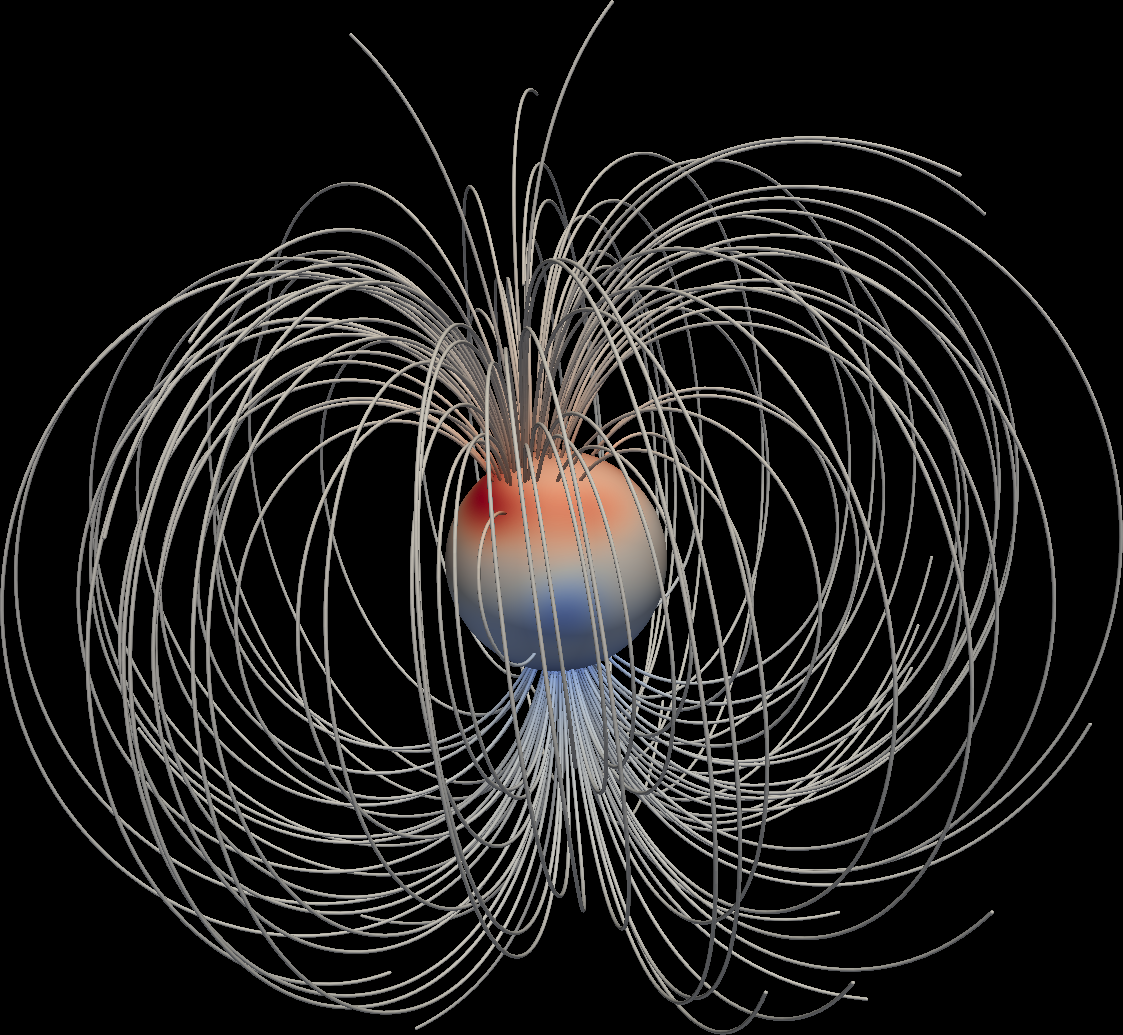}
\caption{3D rendering of Jupiter's magnetic field using Paraview, using a vts file produced by planetMagfields.}
\label{fig:jup3d}
\end{figure}

\hypertarget{magnetic-field-models-used}{%
\subsection{Magnetic field models
used}\label{magnetic-field-models-used}}

\texttt{planetMagfields} currently supports the following magnetic field
models:

\begin{itemize}
\tightlist
\item
  \emph{Mercury} : Anderson et al.
  (\protect\hyperlink{ref-Anderson2012}{2012}) , Thébault et al.
  (\protect\hyperlink{ref-Thebault2018}{2018}), Wardinski et al.
  (\protect\hyperlink{ref-Wardinski2019}{2019})
\item
  \emph{Earth} : The International Geomagnetic Reference Field (IGRF)
  (\protect\hyperlink{ref-IGRF13}{Alken et al., 2021})
\item
  \emph{Jupiter} : The VIP-4 model
  (\protect\hyperlink{ref-Connerney1998}{Connerney et al., 1998}), JRM09
  (\protect\hyperlink{ref-Connerney2018}{Connerney et al., 2018}), JRM33
  (\protect\hyperlink{ref-Connerney2022}{Connerney et al., 2022})
\item
  \emph{Saturn} : Cassini Saturn orbit insertion (SOI)
  (\protect\hyperlink{ref-Burton2009}{Burton et al., 2009}), Cassini11
  (\protect\hyperlink{ref-Dougherty2018}{Dougherty et al., 2018}),
  Cassini11+ (\protect\hyperlink{ref-Cao2020}{Cao et al., 2020})
\item
  \emph{Uranus} : Connerney et al.
  (\protect\hyperlink{ref-Connerney1987}{1987})
\item
  \emph{Neptune} : Connerney et al.
  (\protect\hyperlink{ref-Connerney1991}{1991})
\item
  \emph{Ganymede} : Kivelson et al.
  (\protect\hyperlink{ref-Kivelson2002}{2002})
\end{itemize}

When new magnetic field models become available, either through newly
available data or through reanalysis of existing observations, we will
add them to the current repository, either ourselves or through a
community effort of pull requests.

\hypertarget{jupyter-frontend}{%
\subsection{Jupyter frontend}\label{jupyter-frontend}}

We provide a Jupyter notebook, along with a binder link (
\url{https://mybinder.org/v2/gh/AnkitBarik/planetMagFields/HEAD?labpath=\%2FExploreFieldsInteractively.ipynb}
) that gives interactive access for visualizing the radial magnetic
fields and the corresponding Lowes spectra at various depths, with
different background color options.

\hypertarget{documentation}{%
\section{Documentation}\label{documentation}}

The software has been documented using Sphinx
(\url{https://www.sphinx-doc.org/}) and the documentation is available
here: \url{https://ankitbarik.github.io/planetMagFields/}.

\hypertarget{acknowledgements}{%
\section{Acknowledgements}\label{acknowledgements}}

We thank Thomas Gastine and Jonathan Aurnou for testing the package,
reporting bugs and for feature suggestions. We also thank Maximilian
Arthus Schanner for the \texttt{setup.py} script. We thank Pankaj K.
Mishra and Marshall Styczinski for advising us to get the paper
published in JOSS.

\hypertarget{references}{%
\section*{References}\label{references}}
\addcontentsline{toc}{section}{References}

\hypertarget{refs}{}
\begin{CSLReferences}{1}{0}
\leavevmode\vadjust pre{\hypertarget{ref-Ahrens2005}{}}%
Ahrens, J., Geveci, B., \& Law, C. (2005). 36 - ParaView: An end-user
tool for large-data visualization. In C. D. Hansen \& C. R. Johnson
(Eds.), \emph{Visualization handbook} (pp. 717--731).
Butterworth-Heinemann.
\url{https://doi.org/10.1016/B978-012387582-2/50038-1}

\leavevmode\vadjust pre{\hypertarget{ref-IGRF13}{}}%
Alken, P., Thébault, E., Beggan, C. D., Amit, H., Aubert, J.,
Baerenzung, J., Bondar, T. N., Brown, W. J., Califf, S., Chambodut, A.,
Chulliat, A., Cox, G. A., Finlay, C. C., Fournier, A., Gillet, N.,
Grayver, A., Hammer, M. D., Holschneider, M., Huder, L., \ldots{} Zhou,
B. (2021). {International Geomagnetic Reference Field: the thirteenth
generation}. \emph{Earth, Planets and Space}, \emph{73}(1), 49.
\url{https://doi.org/10.1186/s40623-020-01288-x}

\leavevmode\vadjust pre{\hypertarget{ref-Anderson2012}{}}%
Anderson, B. J., Johnson, C. L., Korth, H., Winslow, R. M., Borovsky, J.
E., Purucker, M. E., Slavin, J. A., Solomon, S. C., Zuber, M. T., \&
McNutt, Jr., Ralph L. (2012). {Low-degree structure in Mercury's
planetary magnetic field}. \emph{Journal of Geophysical Research
(Planets)}, \emph{117}, E00L12.
\url{https://doi.org/10.1029/2012JE004159}

\leavevmode\vadjust pre{\hypertarget{ref-Ayachit2015}{}}%
Ayachit, U. (2015). \emph{The ParaView guide: A parallel visualization
application}. Kitware, Inc. ISBN:~1930934300

\leavevmode\vadjust pre{\hypertarget{ref-Burton2009}{}}%
Burton, M. E., Dougherty, M. K., \& Russell, C. T. (2009). Model of
saturn's internal planetary magnetic field based on cassini
observations. \emph{Planetary and Space Science}, \emph{57}(14),
1706--1713. \url{https://doi.org/10.1016/j.pss.2009.04.008}

\leavevmode\vadjust pre{\hypertarget{ref-Cao2020}{}}%
Cao, H., Dougherty, M. K., Hunt, G. J., Provan, G., Cowley, S. W. H.,
Bunce, E. J., Kellock, S., \& Stevenson, D. J. (2020). The landscape of
saturn's internal magnetic field from the cassini grand finale.
\emph{Icarus}, \emph{344}, 113541.
\url{https://doi.org/10.1016/j.icarus.2019.113541}

\leavevmode\vadjust pre{\hypertarget{ref-visit}{}}%
Childs, H., Brugger, E., Whitlock, B., Meredith, J., Ahern, S., Pugmire,
D., Biagas, K., Miller, M. C., Harrison, C., Weber, G. H., Krishnan, H.,
Fogal, T., Sanderson, A., Garth, C., Bethel, E. W., Camp, D., Rubel, O.,
Durant, M., Favre, J. M., \& Navratil, P. (2012). \emph{{High
Performance Visualization--Enabling Extreme-Scale Scientific Insight}}.
\url{https://doi.org/10.1201/b12985}

\leavevmode\vadjust pre{\hypertarget{ref-Connerney1987}{}}%
Connerney, J. E. P., Acuna, M. H., \& Ness, N. F. (1987). {The magnetic
field of Uranus}. \emph{Journal of Geophysical Research},
\emph{92}(A13), 15329--15336.
\url{https://doi.org/10.1029/JA092iA13p15329}

\leavevmode\vadjust pre{\hypertarget{ref-Connerney1991}{}}%
Connerney, J. E. P., Acuna, M. H., \& Ness, N. F. (1991). {The magnetic
field of Neptune}. \emph{Journal of Geophysical Research}, \emph{96},
19023--19042. \url{https://doi.org/10.1029/91JA01165}

\leavevmode\vadjust pre{\hypertarget{ref-Connerney1998}{}}%
Connerney, J. E. P., Acuña, M. H., Ness, N. F., \& Satoh, T. (1998).
{New models of Jupiter's magnetic field constrained by the Io flux tube
footprint}. \emph{Journal of Geophysical Research}, \emph{103}(A6),
11929--11940. \url{https://doi.org/10.1029/97JA03726}

\leavevmode\vadjust pre{\hypertarget{ref-Connerney2018}{}}%
Connerney, J. E. P., Kotsiaros, S., Oliversen, R. J., Espley, J. R.,
Joergensen, J. L., Joergensen, P. S., Merayo, J. M. G., Herceg, M.,
Bloxham, J., Moore, K. M., Bolton, S. J., \& Levin, S. M. (2018). {A New
Model of Jupiter's Magnetic Field From Juno's First Nine Orbits}.
\emph{Geophysical Research Letters}, \emph{45}(6), 2590--2596.
\url{https://doi.org/10.1002/2018GL077312}

\leavevmode\vadjust pre{\hypertarget{ref-Connerney2022}{}}%
Connerney, J. E. P., Timmins, S., Oliversen, R. J., Espley, J. R.,
Joergensen, J. L., Kotsiaros, S., Joergensen, P. S., Merayo, J. M. G.,
Herceg, M., Bloxham, J., Moore, K. M., Mura, A., Moirano, A., Bolton, S.
J., \& Levin, S. M. (2022). {A New Model of Jupiter's Magnetic Field at
the Completion of Juno's Prime Mission}. \emph{Journal of Geophysical
Research (Planets)}, \emph{127}(2), e07055.
\url{https://doi.org/10.1029/2021JE007055}

\leavevmode\vadjust pre{\hypertarget{ref-Dougherty2018}{}}%
Dougherty, M. K., Cao, H., Khurana, K. K., Hunt, G. J., Provan, G.,
Kellock, S., Burton, M. E., Burk, T. A., Bunce, E. J., Cowley, S. W. H.,
Kivelson, M. G., Russell, C. T., \& Southwood, D. J. (2018). {Saturn's
magnetic field revealed by the Cassini Grand Finale}. \emph{Science},
\emph{362}(6410), aat5434. \url{https://doi.org/10.1126/science.aat5434}

\leavevmode\vadjust pre{\hypertarget{ref-Finlay2020}{}}%
Finlay, C. C., Kloss, C., Olsen, N., Hammer, M. D., Tøffner-Clausen, L.,
Grayver, A., \& Kuvshinov, A. (2020). {The CHAOS-7 geomagnetic field
model and observed changes in the South Atlantic Anomaly}. \emph{Earth,
Planets and Space}, \emph{72}(1), 156.
\url{https://doi.org/10.1186/s40623-020-01252-9}

\leavevmode\vadjust pre{\hypertarget{ref-numpy}{}}%
Harris, C. R., Millman, K. J., Walt, S. J. van der, Gommers, R.,
Virtanen, P., Cournapeau, D., Wieser, E., Taylor, J., Berg, S., Smith,
N. J., Kern, R., Picus, M., Hoyer, S., Kerkwijk, M. H. van, Brett, M.,
Haldane, A., Río, J. F. del, Wiebe, M., Peterson, P., \ldots{} Oliphant,
T. E. (2020). Array programming with {NumPy}. \emph{Nature},
\emph{585}(7825), 357--362.
\url{https://doi.org/10.1038/s41586-020-2649-2}

\leavevmode\vadjust pre{\hypertarget{ref-matplotlib}{}}%
Hunter, J. D. (2007). Matplotlib: A 2D graphics environment.
\emph{Computing in Science \& Engineering}, \emph{9}(3), 90--95.
\url{https://doi.org/10.1109/MCSE.2007.55}

\leavevmode\vadjust pre{\hypertarget{ref-James2024}{}}%
James, M. K., Provan, G., Kamran, A., Wilson, R. J., Vogt, M. F.,
Brennan, M. J., \& Cowley, S. W. H. (2024). \emph{JupiterMag} (Version
v1.3.1). Zenodo. \url{https://doi.org/10.5281/zenodo.10602418}

\leavevmode\vadjust pre{\hypertarget{ref-Jones2011}{}}%
Jones, C. A. (2011). {Planetary Magnetic Fields and Fluid Dynamos}.
\emph{Annual Review of Fluid Mechanics}, \emph{43}(1), 583--614.
\url{https://doi.org/10.1146/annurev-fluid-122109-160727}

\leavevmode\vadjust pre{\hypertarget{ref-Khurana2020}{}}%
Khurana, K. K. (2020). \emph{KMAG - kronian magnetic field model}
(Version 1.0). Zenodo. \url{https://doi.org/10.5281/zenodo.4080294}

\leavevmode\vadjust pre{\hypertarget{ref-Kivelson2002}{}}%
Kivelson, M. G., Khurana, K. K., \& Volwerk, M. (2002). {The Permanent
and Inductive Magnetic Moments of Ganymede}. \emph{Icarus},
\emph{157}(2), 507--522. \url{https://doi.org/10.1006/icar.2002.6834}

\leavevmode\vadjust pre{\hypertarget{ref-Kloss2024}{}}%
Kloss, C. (2024). \emph{Ancklo/ChaosMagPy: ChaosMagPy v0.13} (Version
v0.13). Zenodo. \url{https://doi.org/10.5281/zenodo.10598528}

\leavevmode\vadjust pre{\hypertarget{ref-cartopy}{}}%
Met Office. (2010 - 2015). \emph{Cartopy: A cartographic python library
with a matplotlib interface}. \url{https://scitools.org.uk/cartopy}

\leavevmode\vadjust pre{\hypertarget{ref-Moore2018}{}}%
Moore, K. M., Yadav, R. K., Kulowski, L., Cao, H., Bloxham, J.,
Connerney, J. E. P., Kotsiaros, S., Jørgensen, J. L., Merayo, J. M. G.,
Stevenson, D. J., Bolton, S. J., \& Levin, S. M. (2018). {A complex
dynamo inferred from the hemispheric dichotomy of Jupiter's magnetic
field}. \emph{Nature}, \emph{561}(7721), 76--78.
\url{https://doi.org/10.1038/s41586-018-0468-5}

\leavevmode\vadjust pre{\hypertarget{ref-ipython}{}}%
Pérez, F., \& Granger, B. E. (2007). {IP}ython: A system for interactive
scientific computing. \emph{Computing in Science and Engineering},
\emph{9}(3), 21--29. \url{https://doi.org/10.1109/MCSE.2007.53}

\leavevmode\vadjust pre{\hypertarget{ref-shtns}{}}%
Schaeffer, N. (2013). Efficient spherical harmonic transforms aimed at
pseudospectral numerical simulations. \emph{Geochemistry, Geophysics,
Geosystems}, \emph{14}(3), 751--758.
\url{https://doi.org/10.1002/ggge.20071}

\leavevmode\vadjust pre{\hypertarget{ref-Schubert2011}{}}%
Schubert, G., \& Soderlund, K. M. (2011). {Planetary magnetic fields:
Observations and models}. \emph{Physics of the Earth and Planetary
Interiors}, \emph{187}(3), 92--108.
\url{https://doi.org/10.1016/j.pepi.2011.05.013}

\leavevmode\vadjust pre{\hypertarget{ref-Stanley2014}{}}%
Stanley, S. (2014). Chapter 6 - magnetic field generation in planets. In
T. Spohn, D. Breuer, \& T. V. Johnson (Eds.), \emph{Encyclopedia of the
solar system (third edition)} (Third Edition, pp. 121--136). Elsevier.
\url{https://doi.org/10.1016/B978-0-12-415845-0.00006-2}

\leavevmode\vadjust pre{\hypertarget{ref-Styczinski2024}{}}%
Styczinski, M. J., \& Cochrane, C. J. (2024).
\emph{{coreyjcochrane/PlanetMag: Model updates following publication
peer review}} (Version v1.0.2). Zenodo.
\url{https://doi.org/10.5281/zenodo.10864719}

\leavevmode\vadjust pre{\hypertarget{ref-Thebault2018}{}}%
Thébault, E., Langlais, B., Oliveira, J. S., Amit, H., \& Leclercq, L.
(2018). A time-averaged regional model of the hermean magnetic field.
\emph{Physics of the Earth and Planetary Interiors}, \emph{276},
93--105. \url{https://doi.org/10.1016/j.pepi.2017.07.001}

\leavevmode\vadjust pre{\hypertarget{ref-scipy}{}}%
Virtanen, P., Gommers, R., Oliphant, T. E., Haberland, M., Reddy, T.,
Cournapeau, D., Burovski, E., Peterson, P., Weckesser, W., Bright, J.,
van der Walt, S. J., Brett, M., Wilson, J., Millman, K. J., Mayorov, N.,
Nelson, A. R. J., Jones, E., Kern, R., Larson, E., \ldots{} SciPy 1.0
Contributors. (2020). {{SciPy} 1.0: Fundamental Algorithms for
Scientific Computing in Python}. \emph{Nature Methods}, \emph{17},
261--272. \url{https://doi.org/10.1038/s41592-019-0686-2}

\leavevmode\vadjust pre{\hypertarget{ref-Wardinski2019}{}}%
Wardinski, I., Langlais, B., \& Thébault, E. (2019). {Correlated
Time-Varying Magnetic Fields and the Core Size of Mercury}.
\emph{Journal of Geophysical Research (Planets)}, \emph{124}(8),
2178--2197. \url{https://doi.org/10.1029/2018JE005835}

\leavevmode\vadjust pre{\hypertarget{ref-Wieczorek2018}{}}%
Wieczorek, M. A., \& Meschede, M. (2018). {SHTools: Tools for Working
with Spherical Harmonics}. \emph{Geochemistry, Geophysics, Geosystems},
\emph{19}(8), 2574--2592. \url{https://doi.org/10.1029/2018GC007529}

\leavevmode\vadjust pre{\hypertarget{ref-Wilson2023}{}}%
Wilson, R. J., Vogt, M. F., Provan, G., Kamran, A., James, M. K.,
Brennan, M., \& Cowley, S. W. H. (2023). {Internal and External Jovian
Magnetic Fields: Community Code to Serve the Magnetospheres of the Outer
Planets Community}. \emph{Space Science Reviews}, \emph{219}(1), 15.
\url{https://doi.org/10.1007/s11214-023-00961-3}

\end{CSLReferences}

\end{document}